\documentclass[aps, pre, twocolumn, showpacs, preprintnumbers, amsmath,
amssymb]{revtex4} 

\usepackage[pdftex]{graphicx}	
\usepackage[svgnames,pdftex]{xcolor}
\usepackage{dcolumn}	
\usepackage{bm}			

\usepackage[pdftex, colorlinks]{hyperref}
\hypersetup{%
pdftitle={Influence of Confinement on Dynamical Heterogeneities in
  Dense Colloidal Samples}, 
pdfsubject={Dynamics of confined of colloidal suspensions},
pdfauthor={Kazem V. Edmond, Carolyn R. Nugent, Eric R. Weeks},
pdfkeywords={colloid, confinement, dynamics, heterogeneities, glass
  transition, supercooled},
bookmarksnumbered,
pdfstartview={FitH},
urlcolor=blue,
citecolor=blue,
linkcolor=blue
}%

\begin{document}

\title{Influence of Confinement on Dynamical Heterogeneities in Dense
Colloidal Samples} 
\author{Kazem V.~Edmond}
\altaffiliation{Current address:
Center for Soft Matter Research, Dept. of Physics, New York
University, New York, New York, USA}
\author{Carolyn R.~Nugent}
\altaffiliation{Current address:
Department of Earth and Space Sciences, UCLA, Los Angeles, California, USA}
\author{Eric R.~Weeks}
\affiliation{Physics Department, Emory University, Atlanta, Georgia 30322, USA}

\date{\today}

\begin{abstract}

We study a dense colloidal suspension confined between two
quasiparallel glass plates as a model system for a supercooled
liquid in confined geometries.  We directly observe the
three-dimensional Brownian motion of the colloidal particles
using laser scanning confocal microscopy.  The particles form
dense layers along the walls, but crystallization is avoided as
we use a mixture of two particle sizes.  A normally liquid-like
sample, when confined, exhibits slower diffusive motion.  Particle
rearrangements are spatially heterogeneous, and the shapes of
the rearranging regions are strongly influenced by the layering.
These rearranging regions become more planar upon confinement.
The wall-induced layers and changing character of the spatially
heterogeneous dynamics appear strongly connected to the confinement
induced glassiness.

\end{abstract}


\maketitle

\section{ Introduction }

As a glass-forming liquid is cooled, its viscosity increases smoothly
but dramatically by many orders of magnitude.  The macroscopic
divergence in viscosity is related to the divergence in the
microscopic structural relaxation time, or $\alpha$-relaxation time.
A conceptual explanation is the Adams and Gibbs hypothesis, which
states that the flow in a supercooled liquid involves the cooperative
motion of molecules and that the structural arrest at the glass
transition is due to a divergence of the size of these cooperatively
rearranging regions (CRRs) \cite{Adam1965Temperature}.

Computer simulations and experiments have explored the sizes
and shapes of regions of cooperatively moving molecules as a
liquid's glass transition is approached \cite{Kob1997Dynamical,
Ediger2000Spatially}.  A direct means of probing the dynamic
length scales of glass-forming liquids is by confining them
to smaller volumes, such as within thin films and nanopores.
Confinement can either increase, decrease
or even maintain a material's glass transition temperature $T_G$
\cite{Alcoutlabi2005Effects,roth05}.  Both simulation and
experiment suggest that the effect on $T_G$ depends on the
nature of the interaction between the sample and its confining
boundary \cite{Nemeth1999Freezing, Scheidler2002Cooperative,
Ngai2002Relaxation, sharp03, Goel2008Tuning, Eral2009Influence,
Rice2009Structure}.  Attractive interactions may result in an
increase in $T_G$ whereas repulsive interactions may result in a
decrease \cite{Roth2007Eliminating, Goel2008Tuning}.  Frustration of
structural ordering, via a rough surface for example, can also
play a key role, although this can either cause slower or faster
dynamics \cite{lowen99,Scheidler2002Cooperative,Eral2009Influence}.
Whether or not the restriction of the length scales accessible to
CRRs is responsible for the variation in $T_G$ remains to be seen
due to the inability to directly observe molecular interactions
within glass-forming liquids.

Instead of studying molecular glass-formers, we use
dense colloidal suspensions of sterically-stabilized
micrometer-sized spherical particles.  Colloidal suspensions
have often been used as experimental models of a hard sphere
glass \cite{Pusey1986Phase,vanBlaaderen1995RealSpace}.
We confine our samples within a planar volume formed by two
quasi-parallel solid surfaces \cite{Nugent2007Colloidal},
similar to confined colloids studied by other groups
\cite{sarangapani08,Eral2009Influence,sarangapani11}.  We use
high-speed confocal microscopy to rapidly visualize and
acquire three-dimensional images of the particle positions
\cite{Weeks2000ThreeDimensional,dinsmore01}.
Subsequent image analysis lets us track the individual
particle trajectories, providing an accurate picture
of the cooperatively rearranging groups of particles.
Near the colloidal glass transition ($\phi_g \approx 0.58$
\cite{Pusey1986Phase}), particles rearrange in groups
characterized by a length scale of $\sim$3-6 particle diameters
\cite{Weeks2000ThreeDimensional,weeks07cor}.


\begin{figure}[b]
\centerline{
\includegraphics[width=0.48\textwidth]{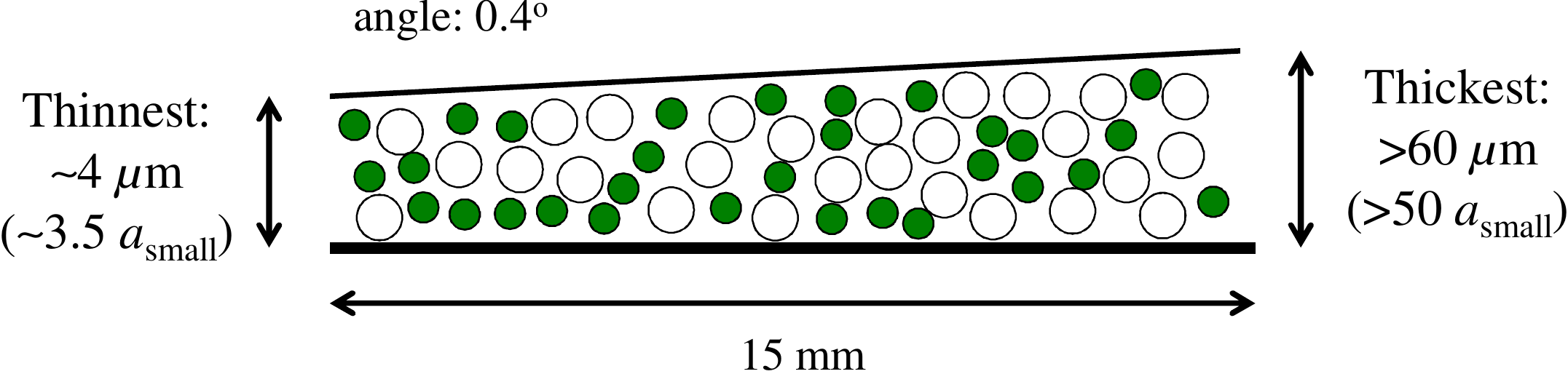}}
\caption{ (Color online) Sketch of sample chamber (not to scale).  The
small particles are $1.18$~$\mu$m in radius and are shaded to indicate 
their fluorescent dye.  The large particles are
$1.55$~$\mu$m in radius and drawn in white to indicate their lack of
dye, making them invisible to the confocal microscope. One of the
boundaries is a coverslip, rather than a glass slide, indicated by the
thinner line. }
\label{wedgecell}
\end{figure}

In this manuscript we further investigate our results from
prior experiments that studied confined samples, as pictured in
Fig.~\ref{wedgecell} \cite{Nugent2007Colloidal}.  Here we focus
specifically on the nature of cooperative rearrangements within
the confined sample and how they relate to the system's increased
glassiness.  In these experiments we found that confinement induces
glassy behavior at concentrations in which the bulk behavior is
still liquid-like.  Here, we show that confining colloidal liquids
within this planar volume results in cooperatively rearranging
groups of particles that are similarly planar shaped.  We find that
the flattening shapes of the cooperatively rearranging groups are
correlated with the overall slowing of the dynamics, suggesting
a connection between confinement, wall-induced structure, and
glassy behavior.

Understanding the effects of confinement on the glass transition
may help us understand the glass transition in the bulk.
Perhaps more importantly, understanding the properties of
confined fluids also has direct relevance with lubrication
\cite{Granick1999Soft}, the flow of liquids through microfluidic
devices \cite{Thompson1992Phase,Goyon2008Spatial}, and the kinetics
of protein folding \cite{Zhou2008Macromolecular}.

\section{ Experimental Details }
\label{methods}

\subsection{Colloidal samples and microscopy}

We use spherical colloidal poly-methyl-methacrylate (PMMA) particles
that are sterically stabilized to prevent interparticle attraction
\cite{Pusey1986Phase, dinsmore01}.  The particles
are suspended in a mixture of solvents, cyclohexylbromide and
cis- and trans-decalin, which matches both their density and index of refraction
\cite{dinsmore01}.  While our sample is similar to
other types of colloidal suspensions that act like hard spheres
\cite{Pusey1986Phase}, the cyclohexylbromide in our solvent mixture 
induces a slight charge on the surfaces of the particles.  Thus, the
particles have a slightly soft repulsive interaction in addition to
their hard sphere core.  To prevent crystallization, which would be 
readily induced by the smooth walls in our thin planar geometry
\cite{Grier1994Microscopic,Dullens2004Reentrant}, we use a binary
mixture of particles with hydrodynamic radii of $a_{\rm small} =
1.18$~$\mu$m and $a_{\rm large} = 1.55$~$\mu$m.  The number ratio is
approximately $N_S/N_L = 3.5$, and the individual volume fractions
are approximately $\phi_S = 0.26, \phi_L = 0.16$, so the total
overall volume fraction is $\phi=0.42 \pm 0.05$.  The uncertainty
of $\phi$ arises from the difficulty in precisely determining
the individual species' particle size, the polydispersity of
particle sizes ($\sim$ $5\%$ for both species), and difficulties
in determining the relative volume fractions of the two species
\cite{poon12}.
A study of a similar colloidal mixture found the glass transition
for bulk samples to be at $\phi_g \approx 0.58$
\cite{narumi11}.

We use laser scanning confocal microscopy to view the sample
\cite{dinsmore01}.  We can
acquire a three-dimensional image of the sample by scanning a
$50 \times 50 \times 20$~$\mu$m$^3$ region (equal to $256 \times
256 \times 100$ pixels).  We use Visitech's ``vt-Eye'' confocal
system which can scan this volume in 2.0~seconds. This is much
faster than the time for particles to diffuse their own diameter,
which is $\sim$100~seconds in our samples.  We acquire sequences
of three-dimensional (3D) confocal images every 2.0~seconds for
up to 45 minutes.  By scanning different locations, we observe
the behavior at different chamber thicknesses ranging from
$\sim$6~$\mu$m to $\sim$19~$\mu$m in addition to the sample's bulk.
Data representing the `bulk' of our sample is acquired from a
20~$\mu$m thick subvolume in the thicker region of the sample
chamber that is over 15~$\mu$m away from the chamber's walls in
order to avoid any boundary effects.

The small particles are dyed with Rhodamine dye
\cite{dinsmore01} and the larger ones
are left undyed.  Thus the data in our results are for the
smaller particles only.  Each image is post-processed to
find particle positions with an accuracy of 0.05~$\mu$m in
$x$ and $y$ (parallel to the walls) and 0.1~$\mu$m in $z$
(perpendicular to the walls, and parallel to the optical axis
of the microscope).  Given that the particles do not move much
between images, we can link the particle positions in time to get
3D trajectories of the particles' motion throughout the sample
volume \cite{dinsmore01,Crocker1996Methods}.

\subsection{Sample chambers}

Our goal is to study our sample with a range of confinement
thicknesses.  Here we focus on ``thin film like'' confinement
between two flat surfaces. We achieve this by constructing a
wedge shaped sample chamber, as shown in Fig.~\ref{wedgecell}.
We build the chamber using a glass slide, a rectangular glass
coverslip, and a narrow piece of a $\sim$60~$\mu$m thick
Mylar film, employing a method similar to the one used by
Refs.~\cite{Murray1998Phases,Fontecha2005Comparative}.  Using
UV-curing epoxy (Norland 68) we attach the Mylar film near one end
of one side of the glass slide so that it runs perpendicular to
the slide's length.  Next, the glass coverslip is laid across the
slide so that one end is raised up by the Mylar film.  Meanwhile,
the coverslip's opposite end is clamped down, ensuring the thinnest
gap size possible.  We seal the sample chamber shut with epoxy,
except for two small air holes; the sample is added via one while
air escapes via the other.  After adding the sample, the two
openings are sealed with epoxy.  The chamber's shape is described
in Fig.~\ref{wedgecell}: a very long chamber with a broad range
of thicknesses.  Due to the Mylar film, the glass surfaces are
not parallel but very slightly angled at $0.4^\circ$ relative to
one another.  Within our field of view, the change in our sample's
thickness due to our sample chamber's slight taper is less than
$0.3$~$\mu$m, which is negligible for all but the thinnest regions.
We do not see any influence of the taper in any of our results
(discussed further below), suggesting it is reasonable to consider
the two boundaries as locally quasi-parallel.  We define $y$
as the direction along which $H$ varies.

\begin{figure}
\centerline{
\includegraphics[width=8.0 cm]{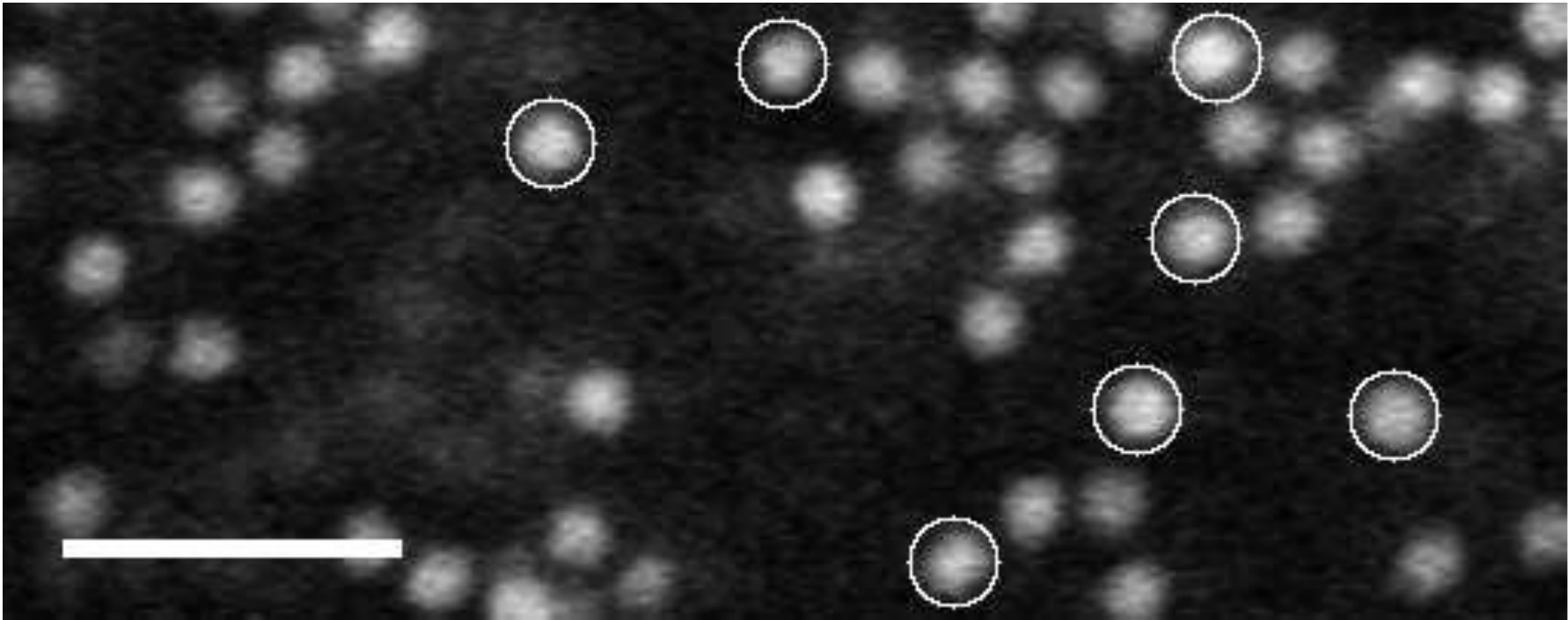}}
\smallskip
\caption{Typical 2D confocal microscope image showing particles
immediately adjacent to one of the chamber walls.  The circled
particles are stuck to the glass, and the others move freely. 
There are also undyed particles also stuck to the
surface, as well as undyed mobile particles, which are not
visible in this confocal image.
The
scale bar indicates 10~$\mu$m. 
}
\label{fover}
\end{figure}

When we fill our slides with sample, a small fraction of
particles stick to the sample chamber's walls.  Typically less
than $20\%$ of the walls' area is coated with stuck particles
\cite{edmond10b}.  The stuck particles are easy to identify as
their apparent motion, due to noise inherent to particle tracking,
is much less than the other particles.  An image showing the
locations of some stuck particles is shown in Fig.~\ref{fover}.
Other observations confirm that both large and small particles
stick to the walls \cite{edmond10b}.  We find that the particles
stick to the surfaces of the glass slides only during the initial
loading of the sample chamber with colloid.  The stuck particles
remain stuck indefinitely, through a van der Waals attraction to
the glass, and are a permanent feature
of the surface.  The mobile particles
do not stick to the sample's glass boundaries over time
-- during the experiments they never are seen to stick, and over
several months the amount of particles stuck to the glass
does not appear to change.  In fact, the mobile particles are
repelled from the glass boundaries by a relatively weak Coulombic
interaction; in other words, during the course of the experiment,
the only particle-wall interaction is a weakly repulsive one.
In a sample of dilute colloids, we observe that the concentration of
particles is low at the wall and approaches the bulk value quickly,
within $0.5 \pm 0.1$~$\mu$m, suggesting that the Debye screening
length is $\approx 0.4$~$\mu$m at most and more likely $\approx
0.2-0.3$~$\mu$m, in agreement with prior observations \cite{royall05}.
The stuck particles are expected to slightly slow adjacent
particles \cite{eral11}, which has been confirmed in our
experimental data \cite{edmond10b}.

\begin{figure}
\centerline{
\includegraphics[width=0.48\textwidth]{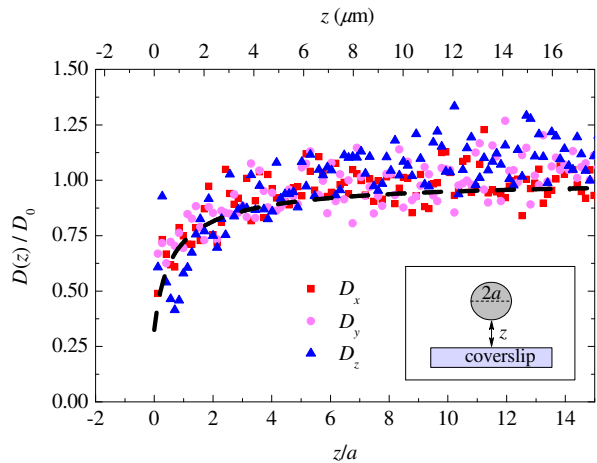}}
\caption{
(Color online)
Measurements of local diffusion constants as a function of the
distance $z$ to the wall, normalized by the small particle radius
$a=1.18$~$\mu$m.  Inset:  sketch indicating that $z=0$
corresponds to the particle touching the wall.
} 
\label{faxen}
\end{figure}

Particles do interact with the wall hydrodynamically.  In the
same dilute suspension, we measure particle mobility near the
glass walls, with measured diffusivity shown in Fig.~\ref{faxen}
as a function of the distance $z$ from the wall.  The behavior
(symbols) is in good agreement with Faxen's Law (dashed line)
\cite{Faxen,svoboda94} which quantifies the hydrodynamic influence
of a planar boundary.  Of course, the hydrodynamic behavior
is modified in confinement approaching quasi-two-dimensional
situations, where the sample chamber thickness $H$ is comparable to
the particle size $2a$ \cite{santanasolano05b,valley07,wonder11}.
We do not consider experiments that are this thin; our
observations all have $H \gtrsim 6a$.  More significantly,
for the larger volume fractions we consider in this work, the
hydrodynamic interaction will be screened by the other particles,
and so will not depend so strongly on the distance from the wall
\cite{Michailidou2009Dynamics,wonder11}.

\section{ Results }

\subsection{ Wall-induced structure}
\label{seclayers}

\begin{figure}
\centerline{
\includegraphics[width=0.48\textwidth]{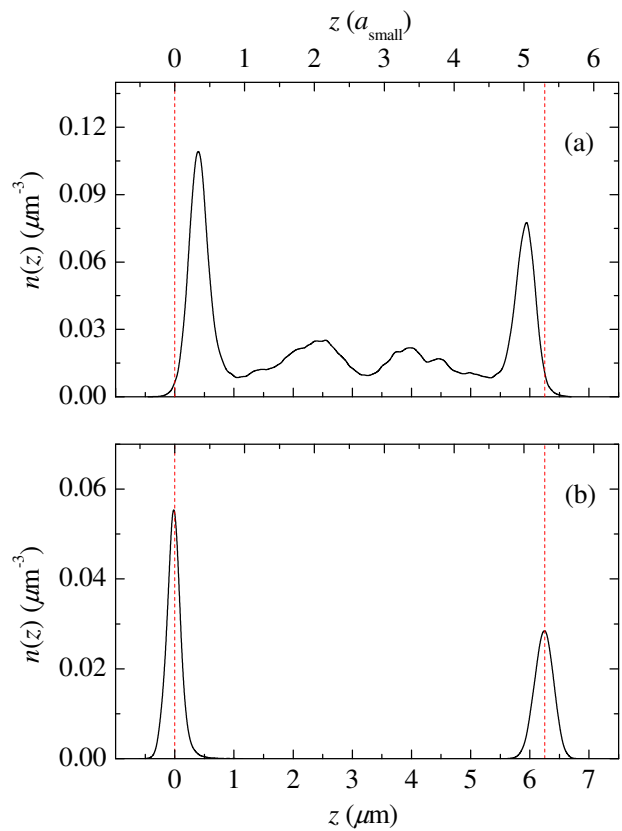}}
\caption{
(Color online)
The number density $n$ as a function of the distance $z$ between the
walls, for the visible (small) particles. 
(a) All mobile particles.
(b) All immobile particles.
The vertical lines in both indicate the position of the centers of the 
visible particles stuck to the walls.  For this data, the distance between 
the two positions is $H=6.25$~$\mu$m, the effective local chamber thickness.
} 
\label{nzstuck}
\end{figure}

We use the positions of the stuck particles to measure the local
thickness of the sample chamber.  To do this we find the number
density $n(z)$ as a function of the distance $z$ between the walls,
shown in Fig.~\ref{nzstuck} for (a) the mobile particles and (b)
the stuck particles.  The maximum of each peak in (b) corresponds
to the approximate position of the \emph{centers} of the
small particles stuck to the sample's walls.  These positions are
marked by the vertical dashed lines in Fig.~\ref{nzstuck}, whose
separation indicate the \emph{effective} local chamber thickness
$H$.  Since only the small particles are visible to the microscope,
the actual thickness is $H+2a_{\rm small}=H+2.36$~$\mu$m.
The mean particle radii are known only to within $\pm0.02$~$\mu$m,
while our uncertainty in their $z$ positions is $0.1$~$\mu$m.
By averaging over tens of stuck particles we can determine $H$
to within $0.01$~$\mu$m.

Figure~\ref{nzstuck}(a) shows layering of particles near the
sample walls, which has been seen in both computer simulations
\cite{Nemeth1999Freezing,Archer2007Dynamics} and experiments
\cite{Murray1998Phases,Dullens2004Reentrant,Eral2009Influence,eral11}.
Comparing Fig.~\ref{nzstuck}(a) to (b) we see that the
boundary layers of the mobile particles are offset from those of the
stuck particles.  The offset is due to Coulombic repulsion between
the glass walls and PMMA particles, and is about 0.4~$\mu$m in all
cases.  Using differential interference contrast (DIC) microscopy,
we confirm that the large particles also form layers, albeit in
positions shifted due to their size.  Our results are qualitatively
in agreement with simulations that studied layering of binary
mixtures of particles near walls
\cite{Desmond2009Random,Mittal2007Confinement}, and
are fairly similar to observations of layering in single-component
colloidal samples
\cite{Dullens2004Reentrant,Eral2009Influence,eral11}.

\begin{figure}
\centerline{
\includegraphics[width=0.48\textwidth]{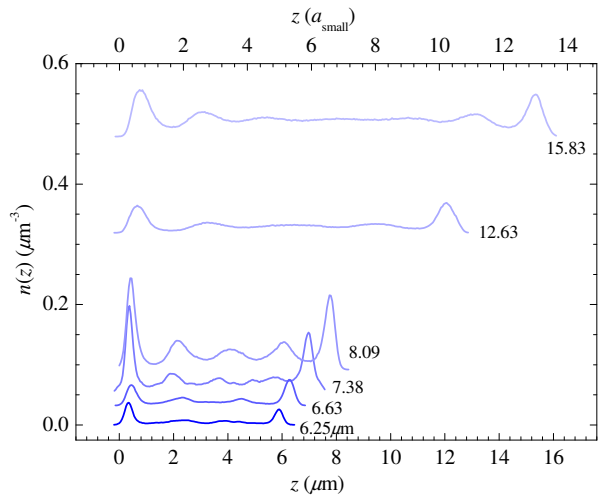}}
\caption{
(Color online)
Number density $n$ as a function of the distance $z$ between the
walls. 
The thickness $H$ is as labeled (in microns).
The values of $H$ in terms of $a_{\rm small}$ are 5.30, 5.62, 6.83,
6.86, 10.7, and 13.4.
The curves are vertically offset for clarity, where the offset
is proportional to $H$.  Where there is an asymmetry in the
height of the $z \approx 0$ peak and the $z \approx H$ peak, it
is due to one wall having more stuck particles on it, thus
decreasing the room available for mobile particles.
} 
\label{manynz}
\end{figure}

Figure~\ref{manynz} displays the way layering changes with $H$.
The peaks of $n(z)$ are tallest and thinnest next to the walls.
Subsequent layers are shorter and wider, presumably as the
correlations between particle positions become diluted through
the presence of two particle sizes \cite{Desmond2009Random}.
Note that we do not see any ``quantization'' effects for particular
values of $H$ \cite{Mittal2008Layering}.  For example, some packing
effects were seen in simulations at $H = 2 m a_{\rm small} + 2 n
a_{\rm large}$ for integer values $m, n$, but these effects are
too subtle to be resolved given the relatively few values of $H$
for which we have experimental data \cite{Desmond2009Random}.

\subsection{ Sample-averaged dynamics }

\begin{figure}
\centerline{
\includegraphics[width=0.48\textwidth]{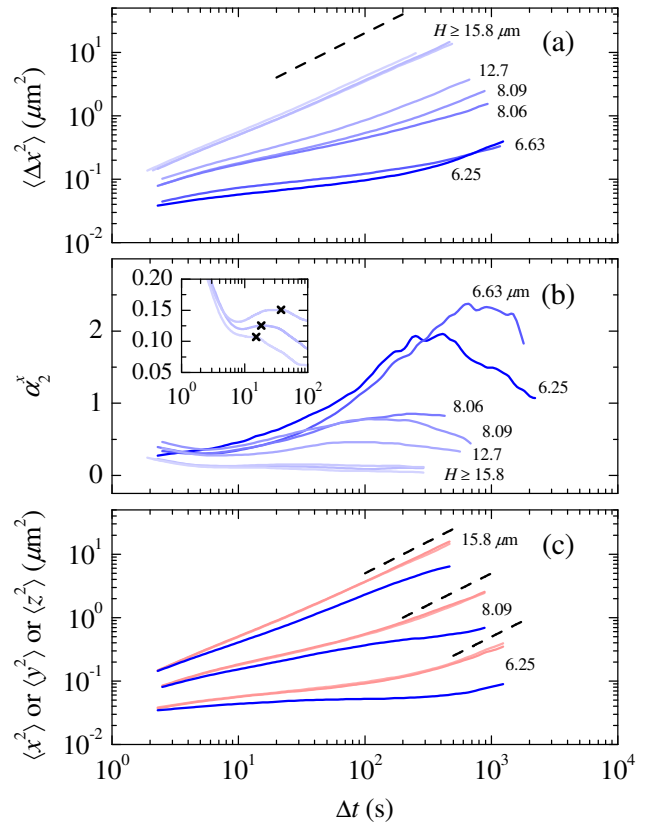}}
\caption{
(Color online)
(a) The mean square displacement for our sample over a range of
thicknesses \cite{Nugent2007Colloidal}.
The dashed line has a slope of 1.0.
(b) Plots of the corresponding non-Gaussian parameter for each
thickness.  The $x$ superscript of $\alpha_2^x$ is to indicate
that the non-Gaussian parameter is only calculated using the $x$
displacements (parallel to the wall, and perpendicular to the
slight gradient in $H$).
The inset is a magnification of the curves for $H \geq 15.8~\mu$m, with
each curve's local maxima labeled, corresponding with $\Delta t^*$
for the data at these thicknesses.
(c) Components of the MSD curves.
Light gray (red) curves
are the $x$- and $y$-components of motion (parallel to the
walls) and the dark gray (blue) are the $z$-component of motion (perpendicular).
}
\label{msd-nong}
\end{figure}

Before we consider the specific influence of the particle layers
on the particle motion, we will quantify the average motion of the
sample.  This is done by calculating the mean square
displacement (MSD) as 
\begin{equation*}
\langle \Delta x^2 \rangle = \langle [x_i(t+\Delta t) - x_i(t)]^2 \rangle_{i,t}
\end{equation*}
where the average is taken over all particles $i$ and all initial
times $t$.  Analogous formulas apply for $\langle y^2 \rangle$
and $\langle z^2 \rangle$.  Figure \ref{msd-nong}(a) shows
that the motion parallel to the walls slows dramatically with
confinement (decreasing $H$, as indicated).  For values less than
$H \approx 16~\mu$m~$\approx 14~a_{\rm small} \approx 10~a_{\rm
large}$ we observe a systematic slowdown.

The change of shape of the curves in Fig.~\ref{msd-nong}(a) suggest
that confinement induces caging dynamics.  This is the inhibited
motion of a particle due to its ``cage'' of neighboring particles
\cite{berne97,bartsch98,doliwa98,Weeks2002Properties,shattuck07}.
At the earliest times ($\Delta t < 1$~s, not shown), particle motion
is diffusive as particles have not moved far enough to encounter the
cage formed by the neighboring particles \cite{tokuyama07}.  As the
particle displacement becomes larger, its motion is impeded by its
neighbors which form the cage, resulting in a greatly decreased
slope of $\langle \Delta x^2 \rangle$ for $\Delta t < 100$~s.
For smaller values of $H$, the decreasing height of $\langle
\Delta x^2 \rangle$ in this range suggests that the cage size
decreases in more confined samples.  This is likely due to the
concentration of particles into the layers (Fig.~\ref{manynz}),
which crowds them within the layers and reduces their cage sizes.
Returning to Fig.~\ref{msd-nong}(a), the upturn at larger
$\Delta t$ for $\langle \Delta x^2 \rangle$ is the result of
cage rearrangements \cite{Weeks2000ThreeDimensional, bartsch98,
Weeks2002Properties, Scheidler2002Cooperative}.  The neighbors
rearrange and this lets the caged particle move to a new position.
The motion of particles at longer lag times is diffusive due to
the uncorrelated cage rearrangements \cite{Weeks2002Properties};
this is not quite seen in our data sets here as the time scales
for this diffusive motion is longer than our observation times.
The results shown are for one volume fraction; our prior work
showed that for larger $\phi$, the onset length scale for the
confinement-induced slow-down increases \cite{Nugent2007Colloidal}.

To contrast the mobility in the parallel and perpendicular
directions, in Fig.~\ref{msd-nong}(c) we plot $\langle x^2 \rangle$
and $\langle z^2 \rangle$ separately for a selection of three
thicknesses.  Not surprisingly, the mobility is less in the $z$
direction (perpendicular to the wall).  Furthermore, the upturn of
the MSD at large $\Delta t$ is barely beginning for the $z$ data.
The contrast between the $x$ and $z$ motion suggests that cage 
rearrangements may favor motions parallel to the walls.

\begin{figure}
\centerline{
\includegraphics[width=0.48\textwidth]{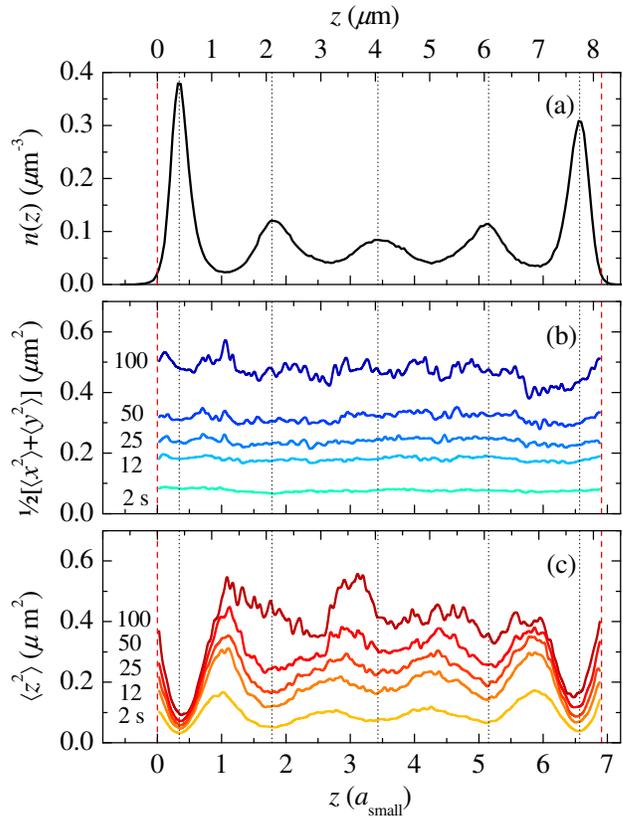}}
\caption{
(Color online)
(a) Particle number-density $n_{\rm small}(z)$ as a function
of distance $z$ across the sample cell.
Additional particles are permanently stuck to the walls of the cell
(not shown) which have centers located at $z=0.00$~$\mu$m and
$z=H=8.06$~$\mu$m, indicated by the vertical dashed lines.
These data correspond to the $H=8.06$~$\mu$m data in
Fig.~\ref{msd-nong}.
(b) Mean square displacement parallel to the walls
($\frac{1}{2}[\langle \Delta x^2 \rangle + \langle \Delta y^2
\rangle]$) and (c) perpendicular to the walls ($\langle \Delta z^2
\rangle$) as a function of the particles' initial positions $z$.
The displacements are calculated using a range of $\Delta t$,
as labeled.  The dotted lines indicate the position of the
number density maximum of each layer in (a) while the dashed
lines correspond to the approximate position of the centers of
the particles stuck to the glass walls.
} 
\label{mobilnz}
\end{figure}

\begin{figure}
\centerline{
\includegraphics[width=0.48\textwidth]{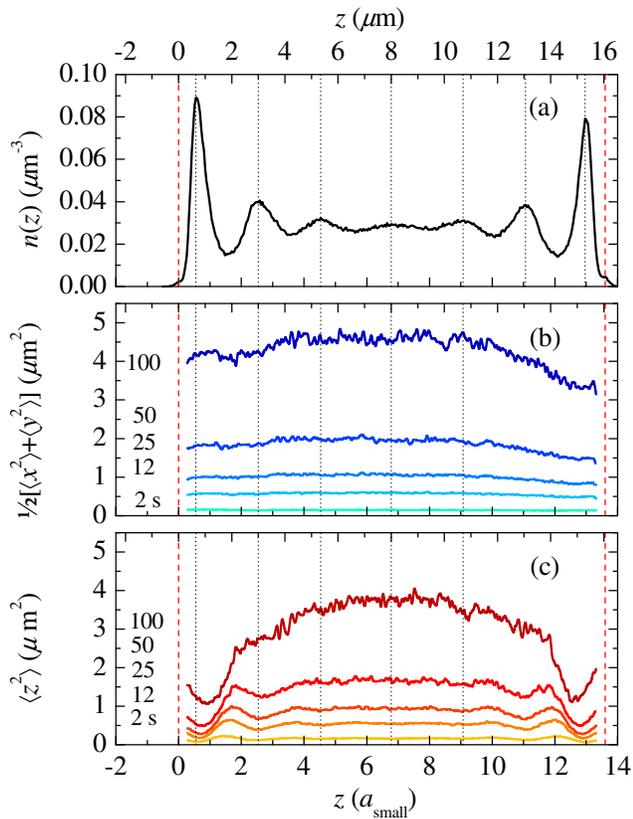}}
\caption{
(Color online)
(a) Particle number-density $n_{\rm small}(z)$ as a function
of distance $z$ across the sample cell.
Additional particles are permanently stuck to the walls of the cell
(not shown) which have centers located at $z=0.00$~$\mu$m and
$z=H=16.0$~$\mu$m, indicated by the vertical dashed lines.
These data correspond to the $H \geq 15.8$~$\mu$m data in
Fig.~\ref{msd-nong}.
(b) Mean square displacement parallel to the walls
($\frac{1}{2}[\langle \Delta x^2 \rangle + \langle \Delta y^2
\rangle]$) and (c) perpendicular to the walls ($\langle \Delta z^2
\rangle$) as a function of the particles' initial positions $z$.
The displacements are calculated using a range of $\Delta t$,
as labeled.  The dotted lines indicate the position of the
number density maximum of each layer in (a) while the dashed
lines correspond to the approximate position of the centers of
the particles stuck to the glass walls.
} 
\label{mobilnz2}
\end{figure}

The MSD curves show an overall slowing down due to confinement,
but obscure the influence of the density layers on the motion.
Figure \ref{mobilnz}(a) shows the number density for one data set.
In panels (b) and (c) we plot the components of the MSD, for fixed
values of $\Delta t$, that are perpendicular and parallel to the
walls.  The dips in $\langle z^2 \rangle$ [Fig.~\ref{mobilnz}(c)]
coincide with the layers in panel (a) and imply that particles
within layers are in a preferred structural configuration and
are less likely to move elsewhere \cite{Nugent2007Colloidal,
Nemeth1999Freezing, Archer2007Dynamics, Mittal2008Layering,
Eral2009Influence}.  Meanwhile, the parallel component of
motion shows no variation with $z$, even for long time scales.
Our observations differ from one prior experiment by Eral {\it et
al.} \cite{Eral2009Influence}.  They found a decreased parallel
mobility near the walls but did not measure perpendicular mobility.
One difference is that they studied a single-component sample
with a polydispersity of 8\%, whereas we study a binary sample.
Another difference is that their experiment had a spatial gradient
in volume fraction due to non-density matched particles (they have
a density difference between solvent and particles of $\Delta
\rho \approx 800$~kg/m$^3$, much larger than our value $\Delta
\rho \approx 0.3$~kg/m$^3$).


Intriguingly, our results shown in Figs.~\ref{manynz},\ref{mobilnz}
look strikingly similar to recent experiments by Wonder,
Lin, and Rice \cite{wonder11}.  They studied a monodisperse
quasi-two-dimensional colloidal system, where particles were limited
to one layer in $z$, and further constricted in $y$ analogous to our
confinement in $z$.  They found that their experimental short-time
diffusion coefficients had a similar qualitative behavior to what
is shown in Fig.~\ref{mobilnz}(b,c) \cite{wonder11}.  They did
not study long-time diffusion coefficients.

Our observed reduced particle mobility perpendicular to the
walls is similar to the observations of Dullens and Kegel, who
studied the first layer of colloidal particles at a smooth glass
surface \cite{Dullens2004Reentrant, Dullens2005Topological}.
In their work, quasi-two-dimensional (q-2D) layers of particles
formed along the surface of a glass slide in a bulk polydisperse
colloidal suspension, just as we observe.  Their wall-based
particles only intermittently exchanged with the bulk particles
\cite{Dullens2005Topological}.  In their q-2D wall layer,
particles exhibited two-dimensional behavior that was fundamentally
distinct from the dynamics of the particles further from the wall.
However, a primary reason for this was that the particles were
fairly monodisperse, and thus could form monodisperse 2D phases
\cite{Dullens2004Reentrant, Dullens2005Topological}.  In our
experiments, the particle layers near the wall become more
pronounced with decreasing $H$, suggesting that these layers
become more q-2D.  However, our samples are binary and the two
particle sizes remain well-mixed even at the walls (confirmed by
DIC microscopy).  While the q-2D nature of our layered particles
may partially explain their slow motion, Fig.~\ref{mobilnz} shows
that slowing is not restricted to these layers alone.  Note that the
hydrodynamic interaction of particles with nearby walls diminishes
as the volume fraction is increased \cite{Michailidou2009Dynamics}.

\begin{figure}
\centerline{
\includegraphics[width=0.48\textwidth]{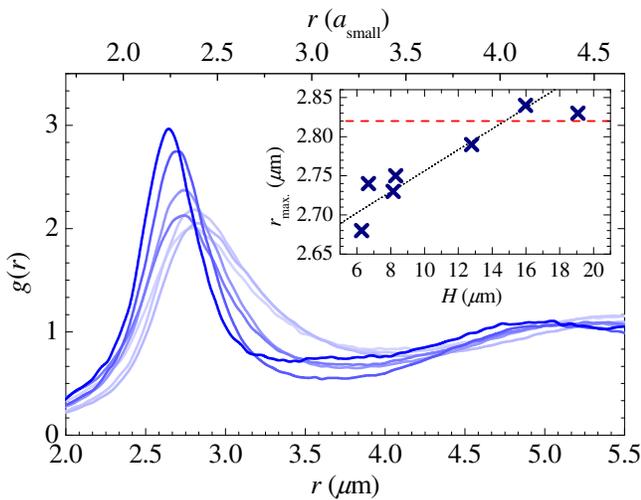}}
\caption{
(Color online)
The pair correlation function $g(r)$ for a range of $H$.
Darker curves correspond with thinner samples.  The curves are
from samples with thickness $H$ equal to 6.25, 6.63, 8.09, 12.6,
15.8, and 18.9~$\mu$m, along with one curve for the sample's bulk
(the lightest color curve).  The inset shows the position of the
first peak as a function of $H$.  The red horizontal dashed line
indicates $r_{\rm max}$ for the bulk sample, while the diagonal
dotted black line is a guide to the eye.
}
\label{gofr}
\end{figure}

One explanation for the slower dynamics might be that the volume
fraction is larger in confinement.  We first consider an observation
from our experiment:  the pair correlation function $g(r)$
changes slightly upon confinement, as shown in Fig.~\ref{gofr}.
This function indicates the likelihood of finding a particle a
distance $r$ away from a reference particle at $r=0$, and so the
first peak position indicates a typical spacing between nearest
neighbor particles.  For ideal hard spheres this first peak
position is always at contact ($r_{\rm max} = 2a_{\rm small}$).
Our particles are slightly charged, so the first peak shifts
to larger values.  The peak is additionally rounded by our finite
resolution and the particle polydispersity \cite{narumi11}.
Given the particle charges, an approximate expectation is that
$\phi \sim r^{-3}_{\rm max}$.  The inset to Fig.~\ref{gofr} shows
that confinement causes $r_{\rm max}$ to shift to lower values,
which would correspond in an increase of $\phi$ from 0.42 to 0.49.
One explanation for this is that, given the layering of particles,
the local volume fraction within a layer is higher than 0.42, and
$g(r)$ is reflecting this local volume fraction [which would be
more heavily weighted in the average used to calculate $g(r)$].
Another possible explanation is that the sample chamber is
effectively thinner than we believe, due to the interactions
between the particles and the walls.  As noted above, the particle
concentration is diminished within 0.4~$\mu$m of the walls.
At the thinnest regions we study, $H \approx 6$~$\mu$m; if the
true value is $\approx 5.2$~$\mu$m, this would correspond to an
increase of $\phi$ from 0.42 to $0.42\times(6/5.2) \approx 0.48$,
consistent with the estimate from the $g(r)$ data.  However,
an effective volume fraction increase from 0.42 to 0.49 seems
unable to explain all of the dramatic slowing of the dynamics
seen in Fig.~\ref{msd-nong}(a).  Consider the data at $\Delta t =
100$~s:  $\langle \Delta x^2 \rangle$ drops by a factor of $\sim 40$
going from the bulk to $H = 6.25$~$\mu$m.  A study of an unconfined
binary suspension similar to ours found a drop of $\sim 3.7$ for
a change of $\phi$ from 0.42 to 0.49 \cite{narumi11}.  Thus we
are left with a factor of ten in additional slowing which is
not due to a possible volume fraction change.  This agrees with
the conclusions of Eral {\it et al.} \cite{Eral2009Influence}.

\subsection{ Defining cooperatively rearranging regions }


The features of our $\langle \Delta x^2 \rangle$ curves
resemble those of bulk supercooled colloidal liquids,
where cage rearrangements play a significant role in
the material's underlying dynamics.  The process of cage
rearrangements leads to a liquid's overall structural relaxation
\cite{Ediger1996Supercooled,Angell2000Ten}.  Adam and Gibbs
were the first to hypothesize the existence of ``cooperatively
rearranging regions'' (CRRs) as a supercooled liquid's means of
increasing its configurational entropy \cite{Adam1965Temperature}.
Prior simulations \cite{Kob1997Dynamical,appignanesi06,keyes11} and
experiments \cite{marcus99,Weeks2000ThreeDimensional,kegel00}
found cooperatively moving regions, defined as groups of neighboring
molecules or particles that collectively rearrange their positions.
The connection between these observations and the CRRs of Adam and
Gibbs is perhaps problematic \cite{dalleferrier07}
and there are other theories that also consider CRRs
\cite{Ediger2000Spatially,hedges09,keyes11}.  Independent of
the theoretical status of CRRs, it is certainly intriguing that
spatially heterogeneous dynamics have been seen in a wide range
of glass-forming systems \cite{Ediger2000Spatially}.  We wish to
see how the character of spatially heterogeneous motions changes
upon confinement.



The precise definition of a cooperatively rearranging region
is open to interpretation.  Our definition is described below,
and is comprised of three key elements: (1) the time scale used
to determine displacements, (2) the threshold for considering a
displacement to be a ``rearrangement,'' and (3) the definition
of which particles are adjacent, such that their motion is
``cooperative.''

\begin{figure}
\centerline{
\includegraphics[width=0.48\textwidth]{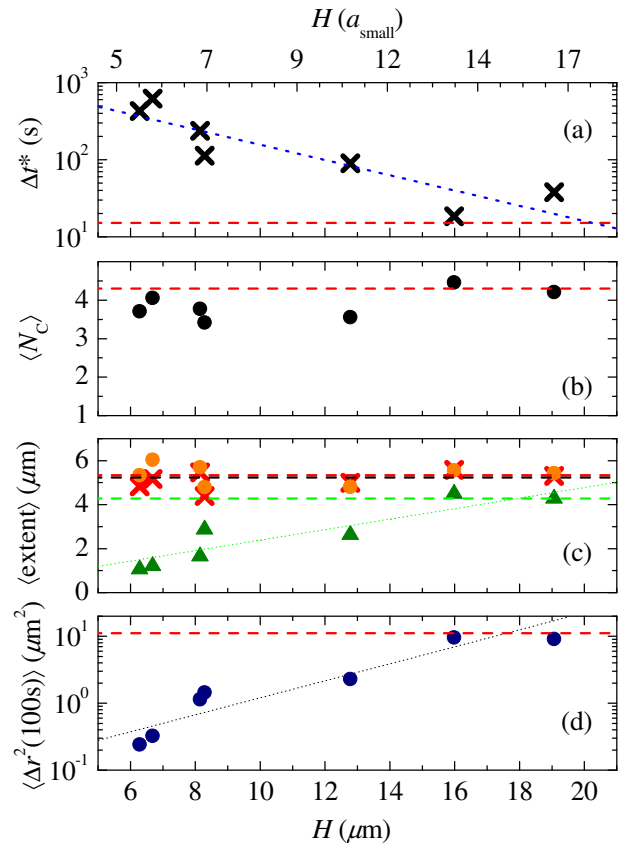}}
\caption{
(Color online)
(a) Values of $\Delta t^*$ that maximize $\alpha_2$ for a range of
thicknesses $H$.
The dotted line is a guide to the eye.
The horizontal dashed line indicates the value of $\Delta t^*$
for the bulk sample.
(b) Plot of the mean number of particles
within cooperatively rearranging regions, as a function of $H$.  Only $N_C
\geq 3$ are considered to avoid trivial rearrangements that consist of
1 or 2 particles.  The upper dashed line indicates $\langle N_C
\rangle$ for the sample's bulk.
(c) Plot of the average extent of 
cooperatively rearranging regions parallel (crosses and circles
correspond to $x$ and $y$, respectively) and
perpendicular (triangles, $z$) to the sample's walls.  The difference
between the $x$ and $y$ data is an indication of the amount of
uncertainty in our data.  The upper and lower horizontal dashed lines
indicate the mean horizontal and perpendicular extent of data from the
sample's bulk, respectively.  In principle these should be the same
(the behavior should be isotropic in the bulk); in practice the lines
may differ due to finite data or anisotropy in the imaging volume
(50~$\mu$m in $x$ and $y$ but only 20~$\mu$m in $z$).
(d) Value of the mean square displacement $\langle \Delta r^2
\rangle$ at the time scale $\Delta t$=100 seconds.  The horizontal dashed line indicates the value of the bulk sample.
}
\label{alphadt}
\end{figure}

We first define the time scale of interest.  Prior work 
found that a good choice is based on the shape of the probability
distribution of displacements.  
Rearranging particles have displacements which are larger than
normal, and thus lie in the tails of the distribution
\cite{kob97,donati98,kegel00,Weeks2000ThreeDimensional,marcus99}.
The size of the distribution tails is quantified
by the non-Gaussian parameter $\alpha_2$,
\begin{equation*}
\alpha_2(\Delta t) = \frac{\langle \Delta x^4 \rangle}{3 \langle \Delta
x^2 \rangle^2} - 1, 
\end{equation*}
from Ref.~\cite{rahman64}.  The maximum of $\alpha_2$ defines
the cage rearrangement time scale $\Delta t^*$.  We plot
$\alpha_2(\Delta t)$ in Fig.~\ref{msd-nong}(b):  both
the maximum value of $\alpha_2$ and the time scale $\Delta t^*$
increase with decreasing $H$, similar to prior observations on a
monodisperse sample \cite{sarangapani11}.  For data from $H \ge 15.8~\mu$m,
the levels of noise at low values of $\Delta t$ manifest as a
false increase of $\alpha_2$, so we ignore this peak.
For $\Delta t > 10$~s there are secondary local maxima of $\alpha_2$
that we consider to be a better determinant of $\Delta t^*$
[see the inset plot of Fig.~\ref{msd-nong}(b)].  We plot $\Delta
t^*$ versus $H$ in Fig.~\ref{alphadt}(a), which decays roughly
exponentially with $H$ until $H \approx 20$~$\mu$m, at which it
reaches the bulk value.  Simply put, as $H$ decreases the
displacement distributions become less Gaussian-like, and the
time scale $\Delta t^*$ for which the distributions are most extreme
grows.





To define the length scale which separates a ``rearranging''
displacement from a ``caged'' displacement, we use
a mobility threshold $\Delta r^*$.  Both experiments
\cite{Weeks2000ThreeDimensional,Keys2007Measurement} and
simulations \cite{Donati1999Spatial} have used a displacement
threshold to define mobility such that over time, some
percentage of the particles have displacements $|\Delta \vec{r}|
\geq \Delta r^*$ \cite{Kob1997Dynamical,Donati1999Spatial},
although at any given time the fraction may not be exactly
this percentage.  Thresholds of the top $5^{\rm th}$ percentile
\cite{Weeks2000ThreeDimensional, Donati1999Spatial}, $8^{\rm th}$
percentile \cite{Weeks2002Properties}, $10^{\rm th}$ percentile
\cite{Keys2007Measurement}, and $20^{\rm th}$ percentile
\cite{Lynch2008Dynamics} have all been used to define $\Delta
r^*$.  From examining distributions of $\Delta x$ and $\Delta y$
for our data for each $H$, we find that the slowest 90\% of the
displacements are well described by a Gaussian distribution, whereas
the top 10\% are more probable than a Gaussian distribution would
predict.  Thus, we define our mobility threshold as the top $10\%$
of the most mobile particles.  Displacements in the $z$-direction,
however, vary significantly with $H$, making their inclusion in
the calculation of our threshold impractical.  As is the case with
prior studies \cite{Weeks2000ThreeDimensional,Kob1997Dynamical},
our choice of $\Delta r^*$ is somewhat arbitrary and our results
are robust to some variation of $\Delta r^*$.

To complete our identification of CRRs we must identify which
highly mobile particles are simultaneously nearest neighbors.
Similar to other work, we define neighbors as those particles
whose separation is less than a cutoff distance set by
the first minimum of the pair correlation function $g(r)$
\cite{Weeks2002Properties}.  Our distributions
of $g(r)$ do not vary substantially with $H$, as shown in
Fig.~\ref{gofr}.  We use the average position of the first
minimum (3.87~$\mu$m) to define particles which are nearest
neighbors.


One problem we face is the selective visibility of the colloidal
particles.  As discussed earlier, only the smaller particles
of our binary suspension are fluorescently labeled meaning that
the larger species of particles are not visible to our confocal
microscope.  Despite this limitation we can still draw some
reasonable conclusions.  For example, in a study of the aging of
a binary colloidal suspension similar to the one studied here,
Lynch {\it et al.} showed that the cooperative dynamics of one
species were similar to that of the other \cite{Lynch2008Dynamics}.
Mobile particles of one species were usually near mobile particles
of the other species.  Therefore, it is reasonable to draw some
conclusions about cooperative motion from the small particles alone.
One other related limitation is that small rearranging particles
may not be nearest neighbors, but may be part of the same CRR,
connected by unseen large particles.  This is beyond our ability
to determine, although it may simply limit the apparent sizes of
CRRs without otherwise changing their character.

\subsection{ Shapes of Cooperatively Rearranging Regions } 

\begin{figure*}
\centering
\includegraphics[width=\textwidth]{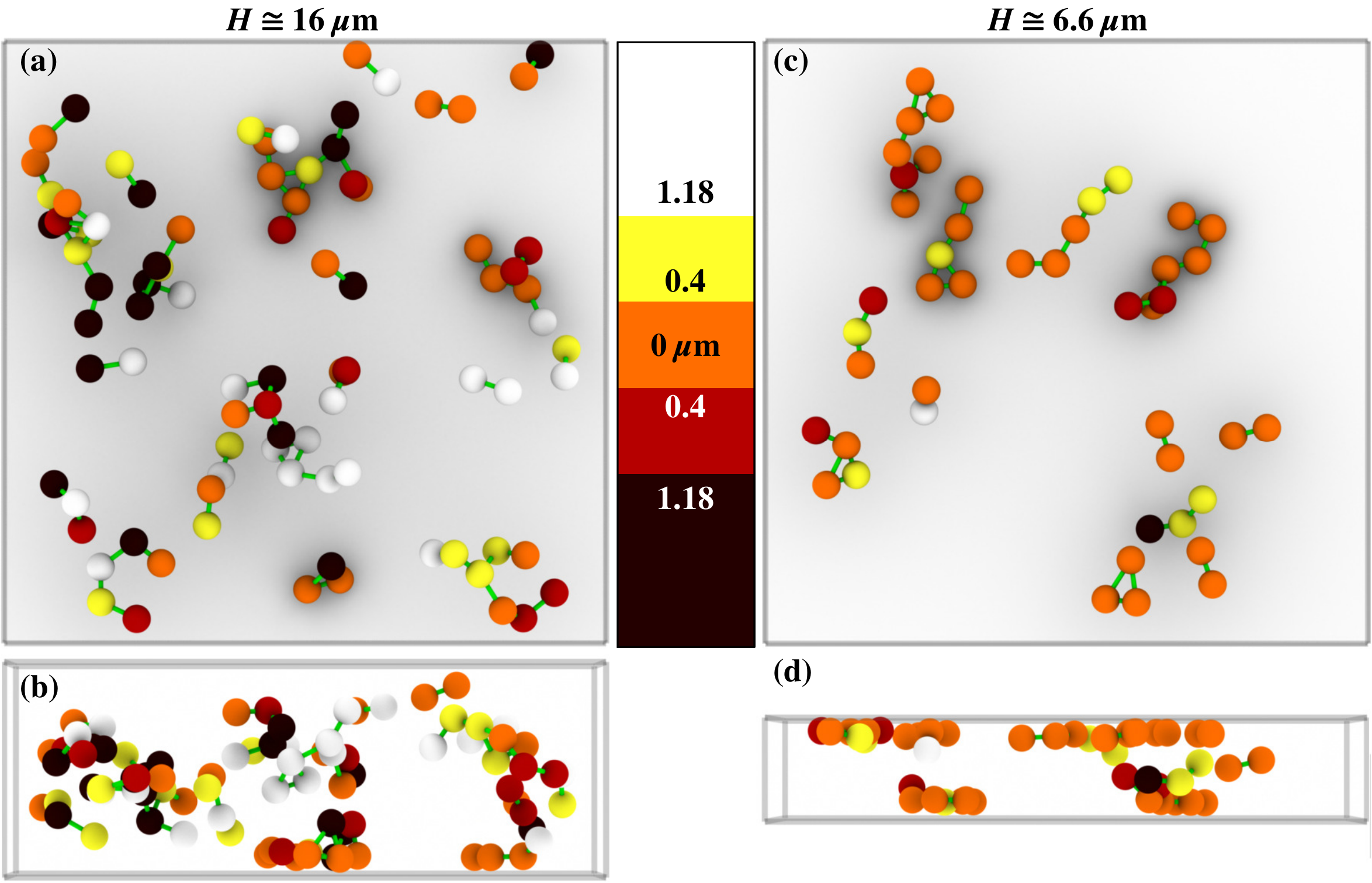}
\caption{
(Color online)
Three-dimensional renderings of the top 10$^{\rm th}$ percentile of the
most mobile particles in two different sample thicknesses.  
The gray (green online) bonds between particles are drawn only to
indicate nearest neighbors and do not imply permanent attachment
between the particles.
Only groups with $N_C~\geq~2$ particles are drawn.
The colors correspond to the magnitude of displacements in the
$z$-direction, normal to the confining boundaries. 
The experiment's field of view, and the effective position of the
confining boundaries, are indicated by the light gray bounding boxes. 
The sample on the left (a, b) has a thickness of $H~=~15.8$~$\mu$m and the one
on the right (c, d) has $H~=~6.63$~$\mu$m.
The top row of images (a, c) view the sample normal to the confining boundaries, while the bottom row (b, d) provide a parallel view.
Black and white indicate displacements of at least $a_{\rm
  small}$ over a $\Delta t~=~23$ s and 250 s for the thicker and
thinner sample respectively.} 

\label{3dlayers}
\end{figure*}

We first visualize these CRRs to develop a qualitative understanding
of their nature.  Figure~\ref{3dlayers} depicts clusters of the top
$10^{\rm th}$ percentile of the most mobile particles in a sample
confined within a plate-spacing of $H = 15.8~\mu$m and $6.63~\mu$m
(panels (a,b) and (c,d) respectively).  
For clarity, bonds have been drawn
between particles that are nearest neighbors, i.e. within a cluster.
For both thicknesses, groups of mobile particles can
be seen.  The size of these mobile clusters in the unconfined sample
is small, as expected for this low volume fraction ($\phi=0.42$)
\cite{Weeks2000ThreeDimensional}.  Despite their small size, these
mobile clusters are the primary means for particle rearrangments
in the sample.  The sample can be considered as composed of the
slowest 90\% particles which are caged at a given moment, and the
rearranging fastest 10\%.  If the nature of the fastest 10\% changes
in confinement -- for example, if those rearrangements occur less
frequently -- then the overall sample diffusivity will decrease.


Confinement induces slower dynamics, and in the
bulk slower dynamics are associated with larger CRRs
\cite{kegel00,Weeks2000ThreeDimensional}.  Perhaps confinement
induces a similar larger size of CRRs \cite{sarangapani11}; but at first glance,
comparing Fig.~\ref{3dlayers} panels (a) and (c) might suggest that
the cluster sizes are smaller upon confinement.  However, recall
that the particles shown are the most mobile 10\%; the thinner
sample has fewer particles in the imaged volume, and thus 10\%
of this smaller number results in fewer mobile particles to show
without necessarily implying that the CRRs are smaller.  To quantify
the size of CRRs we calculate the mean number of particles in a
CRR $N_C$ as a function of $H$, plotted in Fig.~\ref{alphadt}(b).
Figure~\ref{alphadt}(b) shows that CRRs tend to involve roughly
the same number of particles, regardless of thickness.  The mean
CRR size is between 3 and 4 particles, but this is only slightly
larger than the minimum size of 3 particles.  The small size may be
because the bulk sample, with $\phi=0.42$, is liquid-like and only
has small CRRs \cite{Weeks2000ThreeDimensional}.  Alternatively,
as noted above, we cannot see the large particles which are almost
certainly part of CRRs \cite{Lynch2008Dynamics}.  With the data
of Fig.~\ref{alphadt}(b), we cannot say clearly if the CRRs are
larger or smaller upon confinement.  There is a very slight downward
trend in $\langle N_C \rangle$ with decreasing $H$, but this could
be due to poor statistics.  It is possible that the influence of
confinement on the size of CRRs would be clearer in a sample with
a larger value of $\phi$, although such samples are very difficult
to load into our thin sample chambers (as has been noted by others
\cite{haw04,isa09}).  Likely some of the difficulty in loading the
samples is due to their increasing glassiness in confined spaces.
Results from another confocal microscopy experiment on a
monodisperse sample suggested that the length scale for CRRs
grows upon confinement \cite{sarangapani11}.  The difference from
our results may be due to our use of a binary sample.

An alternate way to quantify the size of a CRR is through its
spatial extent.  We define the spatial extent of the CRRs as $x_{\rm
extent} = \text{max}(x_i) - \text{min}(x_i)$, where $i$ ranges
over all particles within a given cluster of mobile particles.
Similar definitions apply for the $y$ and $z$ directions.  We plot
the mean CRR extent in the $x$, $y$ and $z$ directions separately
in Fig.~\ref{alphadt}(c).  We find that the CRRs maintain a
constant size in the direction parallel to the walls.  However,
the amount of distance that the CRRs extend in the direction
perpendicular to the walls is significantly smaller than $H$,
and decreases as $H$ decreases.  In the $z$ direction, then,
clusters are smaller, perhaps trivially because CRRs have to
fit into a thinner sample chamber.  In the $x$ and $y$ direction,
clusters maintain a constant size with confinement.  Comparing this
result with the $\langle N_c \rangle$ data of Fig.~\ref{alphadt}(b)
suggests that the CRRs are becoming more compact in $z$ with the
same number of particles.  This suggests that perhaps they are
fractal in the bulk with a fractal dimension larger than 2 (as seen
previously in Ref.~\cite{Weeks2000ThreeDimensional}) and become
more planar upon confinement (fractal dimension approaching 2).

The onset of flatter or more planar CRRs coincides with the
sample's overall slowing.  In Fig.~\ref{alphadt}(d) we plot the
MSD values from Fig.~\ref{msd-nong}(a) for $\Delta t = 100~$s
against the corresponding range of $H$.  We observe that the
MSD values of Fig.~\ref{alphadt}(d) begin to deviate from those
of the bulk, indicated by the horizontal dashed line in (d),
at approximately the same $H$ that the $z$-extents of the CRRs
first begin to flatten relative to the $z$-extent from the bulk,
the horizontal dashed line in (c).  This is the strongest
evidence linking the changing CRRs to the slowing dynamics.  The
overall concept is that confinement modifies the structure from
that of the bulk, and this changed structure leads to slower
dynamics \cite{Mittal2007Confinement}.

\begin{figure}
\centering
\includegraphics[width=0.48\textwidth]{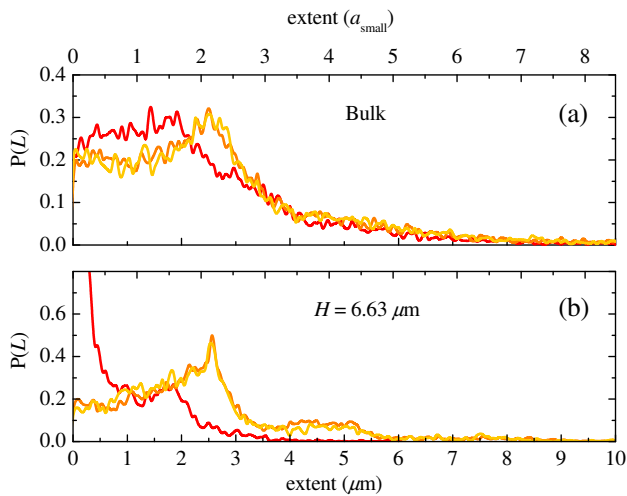}
\caption{
(Color online)
Extent of mobile groups of particles.  Dark and light gray (red and 
orange online) indicate the perpendicular and parallel
extents respectively.  CRRs in the bulk (a) of the sample are more
isotropic in shape than when confined (b) to $H~=~6.63$~$\mu$m.
Note the difference in vertical scales used by the two plots.  Only CRRs
consisting of at least 3 particles are considered in the data shown.
} 
\label{extent}
\end{figure}

We can more carefully quantify the shapes of the CRRs by considering
the probability distributions of the extents in the three
directions.  These distributions are shown in Fig.~\ref{extent} for
a bulk sample (panel a) and a confined sample (panel b).  In the
unconfined sample the probabilities of the extent in the $x$,
$y$, and $z$ directions are approximately the same, as should be
expected; these CRRs are spatially isotropic.  Differences in the
$z$ are most likely due to minor particle position errors which are
larger in $z$, as discussed in Sec.~\ref{methods}.  In contrast with
Fig.~\ref{extent}(a), Fig.~\ref{extent}(b) shows that the extents of
CRRs in confinement have a very different probability distribution.
The extent in $z$ is nearly zero for a majority of CRRs [red
(dark gray) curve in Fig.~\ref{extent}(b)]; these are planar CRRs
and are overwhelmingly more probable than in the unconfined case.
A small subset of confined CRRs do extend into the $z$-direction
by one to two particle diameters.  The clusters of rearranging
particles along the walls in Fig.~\ref{3dlayers}(c, d) seem to be
the most planar in shape.  Thus we are led to conclude that the
CRRs in the confined cases are qualitatively different than those
of the unconfined sample.



\subsection{ Details of Rearrangements }

We next investigate the behavior of particles within CRRs.
In Fig.~\ref{3dlayers} the particles are colored in correspondence
with their amount of perpendicular motion, as shown in the key.
In the confined situation mobile particles displace horizontally
more frequently than otherwise, as suggested by the greater number
of orange (medium gray) particles in Fig.~\ref{3dlayers}(c, d),
This makes sense:  a rearrangement consisting of particles within
a single layer does not require the particles to move vertically
for the rearrangement to occur.  Occasionally we do see particles
which jump between layers or even swap between layers; one example
is near the bottom right corner of Fig.~\ref{3dlayers}(c).

\begin{figure}
\centering
\includegraphics[width=0.48\textwidth]{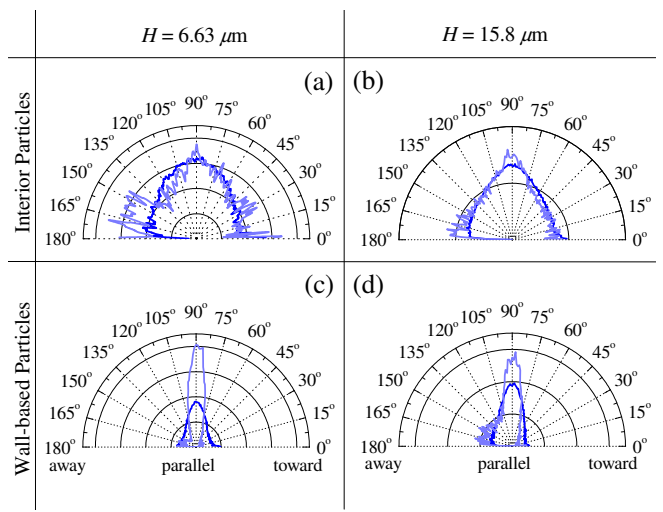}
\caption{
(Color online)
Polar plots of the probability distributions
of the directions of particle displacements for $H=6.63$ and
$H=15.8$~$\mu$m as indicated.  (a,b) Data for the two
thicknesses, considering only particles away from the walls.
(c,d) Data for particles in the layers immediately adjacent to a
wall.  The displacements from one wall are reversed, so that
$180^\circ$ always means motion away from the nearest wall.  In
all panels, the
light blue curve (light gray) is the distribution for the most
mobile 10\% of the particles, while the dark blue curve (dark gray)
is the distribution for all particles.
Displacements are measured over $\Delta t = 250$~s and
$\Delta t = 23$~s for the $H=6.63$~$\mu$m and $H=15.8$~$\mu$m
data, respectively. 
} 
\label{dirdist}
\end{figure}

To compare the amount of parallel versus perpendicular
displacements, we calculate the directions of motion for all
particles and then repeat the comparison for different confinement
thicknesses.  Using a spherical coordinate system we determine
the polar angle of a given particle displacement.  The polar
angle $\theta$ spans a range from $0^\circ$ to $180^\circ$,
which correspond to motion toward or away from the nearest sample
chamber wall, respectively.  That is, we exploit the symmetry
between the two walls.
We first compute the polar angle
$\theta$
relative to the $+z$ axis, and then use $180^\circ - \theta$ for
the data in the lower half of the sample chamber.  Comparing the
data separately for the top and bottom half, we find no
difference in the results.
For isotropic motion, the distribution
of $\theta$ is proportional to $\sin \theta$, so we divide our
measured histograms by $\sin \theta$ to remove this dependence.
The distributions are plotted in polar coordinates, shown in
Fig.~\ref{dirdist}, for thicknesses of $H=6.63$~$\mu$m and $H =
15.8$~$\mu$m.  The dark curves are for all particles,
and the light curves are for the top $10^{\rm th}$
percentile of displacements, providing insight into the directions
that tend to permit higher mobility.  The top panels show the
motion of the particles in the interior of the sample, and the
bottom panels show the motion of the particles immediately adjacent
to the walls.

In both the $15.8$~$\mu$m and $6.63$~$\mu$m samples, the particles
in the outer layers along the walls tend to move parallel to
them ($\theta \approx 90^\circ$) rather than perpendicularly
[Fig.~\ref{dirdist}(c, d)].  The effect is even more pronounced
for the fastest particles, whose distribution suggests that fast
particles move almost exclusively along the walls.  This agrees
with our observations from Fig.~\ref{3dlayers}(b, d), where
the particles layered along the walls are almost all orange
(medium gray), indicating they are moving primarily horizontally.
The distributions in Fig.~\ref{dirdist}(c, d) do show some data at
$\theta=180^\circ$, indicating that some particles move away from
the walls, and less data at $\theta=0^\circ$, indicating that some
particles make slight motions toward the walls.

The situation changes markedly for the inner layers
[Fig.~\ref{dirdist}(a, b)].  Considering only the full distribution
of all particles we see that the displacements are more isotropic,
although there is still a slight bias in the $\theta \approx
90^\circ$ direction.  The distribution of directions for the
most mobile interior particles is similar.  There are bumps in these
distributions near $\theta = 0^\circ$ and $180^\circ$, which
suggests that particles that move in $z$ have a slight increased
probability to make large motions in $z$, hopping between layers.

Overall, the particle dynamics in the thicker region are far
more isotropic than the ones from the confined region [compare
Fig.~\ref{dirdist}(a, c) with (b, d)].  In the $H=15.8$~$\mu$m
case, there are appreciable signs of anisotropic behavior only
along the walls.

\section{ Conclusion }


The smooth quasi-parallel walls confining our sample induce the
formation of density layers within the colloidal sample's volume.
The most dense layers form along the sample chamber's glass
surfaces, as shown in Fig.~\ref{manynz} and also observed in
other experimental work that used single-component colloidal
samples \cite{Dullens2004Reentrant,Eral2009Influence}.
The structural inhomogeneities induced by the density layers
result in corresponding inhomogeneities in the system's dynamics,
as described by the plots in Fig.~\ref{mobilnz}.  Particles move
most easily within their layer, but this is still slower than they
would move in unconfined samples.  The layered particles cooperatively
rearrange within the layer but rarely with adjacent layers; the
cooperative rearrangements occur in more planar-shaped groups of
particles.  Given that even in unconfined samples, particles need
to move cooperatively if they wish to have large displacements, the
change in the character of the cooperatively rearranging regions
seems to explain the slowing dynamics.  In short, the thickness
at which we begin to observe the slowing in the sample's average
dynamics corresponds with the confinement length scale at which
cooperatively rearranging regions begin to become planar in shape
[Figs.~\ref{alphadt}(c) and (d)].  Our prior work suggests that the
observed increase in rearrangement time scales and the thickness
at which these regions begin to flatten will both grow with higher
volume fractions \cite{Nugent2007Colloidal}.

It is likely if the walls were roughened, the results might change.
Simulations \cite{Nemeth1999Freezing, Scheidler2002Cooperative,
Goel2008Tuning} and experiments \cite{Eral2009Influence,eral11}
showed that behavior is often glassier with rough walls.  With rough
walls, layering is greatly diminished or prevented entirely, or
perhaps becomes more subtle.  For example, particles might form a
corrugated layer wrapping around the local wall texture.  This could
then lead to other shapes for the cooperatively rearranging regions;
the main point being that structure that departs from the bulk
results in slower dynamics \cite{Mittal2007Confinement}.

\section{ Acknowledgments }

We thank M.H.G.~Duits, H.B.~Eral, and G.L.~Hunter for helpful
discussions.  Funding for this work was provided by the National
Science Foundation (DMR-0804174).

\end{document}